\documentclass[a4paper, 11pt]{article}
\usepackage{jcappub}
\pdfoutput=1
\usepackage{graphicx}
\usepackage{epsfig}

\newcommand{\ntu}{{Graduate Institute of Astrophysics, National Taiwan University, No.1, Sec.4, Roosevelt Rd, Taipei 10617, Taiwan}}

\title{Large-scale acoustic peak shift in the Planck CMB data}

\author[1, 2]{Ming-Feng Ho}
\author[2]{Lung-Yih Chiang}

\affiliation[1]{\ntu}
\affiliation[2]{\ntu}

\emailAdd{mho026@ucr.edu}
\emailAdd{lychiang@asiaa.sinica.edu.tw}

\abstract{
    Cosmic microwave background (CMB) temperature anisotropies encode the history of the universe, which manifest itself in the angular power spectrum. 
    We test the angular power spectra of small patches from the ESA \textit{Planck} data. 
    Known variations in the power spectra from small patches reveal informative details such as the gravitational lensing and the Doppler boosting effect. 
    We compute the relative shifts of power spectra via comparing patches selected randomly from the CMB. 
    We visualize the relative shifts on a full-sky HEALPix grid (a feature map) and analyze the statistical properties on the full-sky map.
    We find the regions contain the Cold Spot and the Draco supervoid have large relative shifts to large scales.
    We also find a dipole on the generated feature map comparing with simulations.
    We discuss possible ways to resolve this dipole on the feature map, including foregrounds, solar dipole systematics, and the uncertainty in our method.
}


\begin{document}

\maketitle
\flushbottom

\section{Introduction}
Temperature anisotropies of the CMB provided an image of the universe when it was just around 380\,000 years after the Big Bang. They contain a wealth of information about the universe, in particular, the acoustic peaks in the angular power spectrum, a manifestation of compression and rarefaction of the photon-baryon plasma before recombination \cite{Hu2002,Dodelson2003216}. With the measurement of the CMB by NASA WMAP and ESA \textit{Planck} missions, a flat model with cosmological constant and cold dark matter ($\Lambda$CDM model) has emerged as the accepted concordance model for our universe. There are, nevertheless, some anomalies against the model found in WMAP data still persisting in \textit{Planck} data, such as the alignment between CMB quadrupole and octopole and the Cold Spot \cite{wmap7yanomalies,Planck2015xvi}.

Analyses on CMB are usually performed on full-sky maps, but one can do so on small patches too. Acoustic peak measurement from a partial sky not only give us a glimpse of cosmological implication but also shed light on the encoded details inside the patch itself for some CMB secondary effect such as the gravitational lensing \cite{Carron2017}, and the Doppler boosting effect \cite{Planck2013xvii}, both known to shift the underlying power spectrum \cite{Lewis2006, Notari2014, Carron2017}. Also, acoustic peak positions in harmonic space are related to scales; thus, any significant statistical deviation in peak positions of small patches has some physical mechanism attributed to it. 

Due to the importance of the acoustic peaks, the statistics of peak positions of small patches from \textit{Planck} CMB maps was first tested in \cite{Chiang2018}, and it was found the patch covering the famed Cold Spot \cite{Vielva2004, Planck2015xvi}, an area near the Eridanus constellation where the temperature is significantly lower than the Gaussian theory predicts, has significantly large peak shifts relative to the mean. Combined with the significantly low temperature of the Cold Spot, which is also surrounded by surprisingly large underdense regions \cite{Finelli2016}, it was suggested the cause of the anomalies mentioned above ($p=1.1\times 10^{-6}$) is some mysterious energy stretches the space in the transverse direction \cite{Chiang2018}. 

\section{Data Processing}
We followed the previous study in \cite{Chiang2018} to test the full-sky peak shift of small patches in ESA \textit{Planck} Legacy Archive PR2 CMB maps. 
The four \textit{Planck} CMB maps: SMICA, NILC, SEVEM, and COMMANDER maps are derived from different foreground cleaning methods \cite{Planck2015ix}, 
and we used COMMANDER map as the major testing one because it contains much fewer invalid pixels along the Galactic plane, while the other 3 have either artificial spots or significant foreground residuals.

To extract the individual angular power spectrum from small square patches, 
we used a flat-sky approximation to convert the power spectrum $S_k$ obtained via 2D fast Fourier transform (FFT) with the scaling relation \cite{Chiang2012} between Fourier and spherical harmonic transformation

\begin{equation}
    C_{\ell\equiv 2\pi k/L}=  L^2 S_k,     
\end{equation}
where $L$ is the size of the patch. 

Flat-sky approximation enabled us to speed up the computation of angular power spectrum computation, so sampling a large number of patches was possible. 
We also followed the pipeline of the cross-spectrum method used in \cite{Chiang2018}.
The cross-spectrum method used the half-ring maps of CMB and allowed us to eliminate the uncorrelated noise (and its residual \cite{Chiang2011}). 
We also subtracted the extra power induced from non-periodic boundary $ C_\ell ^{\rm NPB} \simeq A \ell ^ {-2.20}$, 
which is described in the data processing pipeline of \cite{Chiang2018}.
Here $A$ is the coefficient varying with different sizes of patches listed in Table 1. 
Finally, we de-convolved the power spectra with a beam transfer function of FWHM 5 arcmin. 

\section{Finding Peak Shifts of Patches}
\begin{figure}
\centering
    \includegraphics[width=\columnwidth]{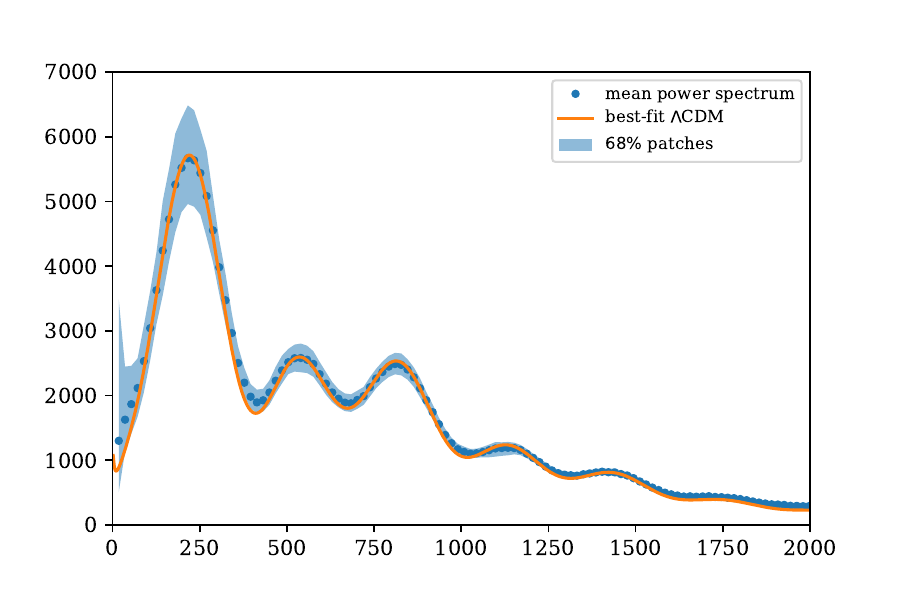}
    \caption{Mean power spectrum of all 30\,720 patches we selected to produce our peak-shift feature map. Blue points are from the mean power spectrum, 
    and the shaded area is the 68 percentile of the power spectra from all 30\,720 patches. 
    The underlying orange curve is the best-fit $\Lambda$CDM model from \textit{Planck}.}
    \label{fig:Mean-Power-Spectrum}
\end{figure}

In our previous work, \cite{Chiang2018}, the patches are selected without overlapping.
However, it is uncertain how much the peak-shift values will vary depending on the selection of patches.
Therefore, here we randomly sampled 30\,720 patches from the COMMANDER half-ring maps, each $20\times20$ deg$^2$.
We furthermore computed uncertainties due to the selections of patches in a given local region.
For each power spectrum $D_\ell$, we fitted with four Gaussian functions.
We used first three peak positions only ${\ell}\equiv(\ell^{(1)}, \ell^{(2)},\ell^{(3)})$ because the amplitudes of the 4th onward are low.
The mean power spectrum of all 30\,720 patches is shown in Fig~\ref{fig:Mean-Power-Spectrum}.

Peak positions were estimated by fitting power spectrum $D_\ell\equiv  \ell (\ell+1) C_\ell/2 \pi$ with the sum of four Gaussian functions from $\ell = 54$ to 1\,350 for $20^2$ deg$^2$ patches (see Table 1 for fitting range for other sizes.\footnote{ Due to the increase of difficulty of fitting in smaller sizes, the available fitting range will shrink in the power spectra with smaller patch sizes }).
We used Levenberg-Marquardt method \cite{More1977} to fit the peak positions \cite{Page2003, Planck2015xi},
which was one of the standard ways applied by WMAP and \textit{Planck}.\footnote{We, in general, applied Levenberg-Marquardt algorithm and used other algorithms such as L-BFGS-B \cite{Zhu1997}, Powell \cite{Powell1964}, TNC \cite{Nash1984}, COBYLA \cite{Powell1998}, and Nelder-Mead \cite{Nelder1965}) for some misfitted patches from the results of Levenberg-Marquardt fitting, though we would like to emphasize that the resultant feature maps change little regardless of re-fitting using other algorithms}

Here we defined statistical quantities of peak shifts we were interested in analyzing.
The ``distance'' of the peak positions to their mean $|{\ell}-\bar{\ell}|$ was used to describe the peak shifts in \cite{Chiang2018},

\begin{equation}
    D =  \sqrt{ \sum\limits_{j = 1}^3 ({\ell}^{(j)} - \bar \ell^{(j)})^2 }.
\end{equation}

The sum of the peak shifts, $I$, in a patch describes the tendency of the first three peaks in a patch power spectrum shifting toward large scales or small scales.

\begin{equation}
    I = \sum\limits_{j = 1}^3 ({\ell}^{(j)} - \bar \ell^{(j)}),
\end{equation}
where $\bar \ell^{(j)}$, $j=1,2,3$ are the means of first three peak positions, respectively. 
Total peak shifts $I$ characterizes the synchronicity of peak shifts in the angular power spectra, and the sign indicates a general tendency of the power spectrum is shifting towards large scales (small $\ell$) for negative $I$ or smaller scales (large $\ell$) for positive $I$.

\section{Feature Maps}

A better way to visualize the tendency of peak shifts on a full-sky is by using a HEALPix grid to smooth the peak shift features on the sphere.
Since the centers of the 30\,720 patches are randomly selected on the sphere,
we then applied low-resolution HEALPix map \cite{Gorski2005} with $N_{\rm side}=16$ (total pixel number 3\,072 with pixel size $\simeq 13.43$ deg$^2$) 
so that each HEALPix pixel can bin 10 patches on average \cite{Kovacs2015}.
For each pixel on the HEALPix grid, we define that in each pixel $\mathcal{I}\equiv \langle I \rangle$ is the mean from all the sampled patches whose centers fall inside the pixel.

The feature map-making process is the same as making a convolutional feature map in computer vision: we slide the convolutional filter (patch)\footnote{patch and filter are interchangeable in this article.} through the whole sphere and record the output of the filter in the feature map (with average pooling). 
Just like the sliding-window method used in convolution, 
the feature in each HEALPix pixel of our feature map is a summarized quantity of $20^2$ deg$^2$ patches.
Each pixel on the feature map is an average feature summarized by nearby patches which have patch centers inside the pixel.
Since $20^2$ deg$^2$ size is larger than the HEALPix pixel size which we chose ($\simeq 13.43$ deg$^2$),  the feature map intrinsically smoothed the feature across the sky.

Ideally, if we shrink the size of patches to smaller, we may have a finer feature map by using a higher resolution HEALPix grids.
However, the fittings of angular power spectra would also become harder since we are getting less information from CMB.
We have explored the other two smaller sizes of feature maps ($15^2$deg$^2$ \& $18^2$deg$^2$), 
and we found that the final feature maps would not change a lot.

\begin{table}
    \centering
    \begin{tabular}{clclclcl}
        \hline
        patch size& $15^2$deg$^2$ & $18^2$deg$^2$ & $20^2$deg$^2$  \\

        coeff. A &  707.17& 509.17 & 492.14  \\
        fitting range in $\ell$ & [96,1350] & [80,1350] & [54,1350] \\
        \hline
    \end{tabular}
    \caption{The coefficient $A$ in non-periodic boundary correction and the fitting ranges for various sizes of square patches.}
    \label{tab:A}
\end{table}

\begin{figure}
    \centering
    \includegraphics[width=\columnwidth]{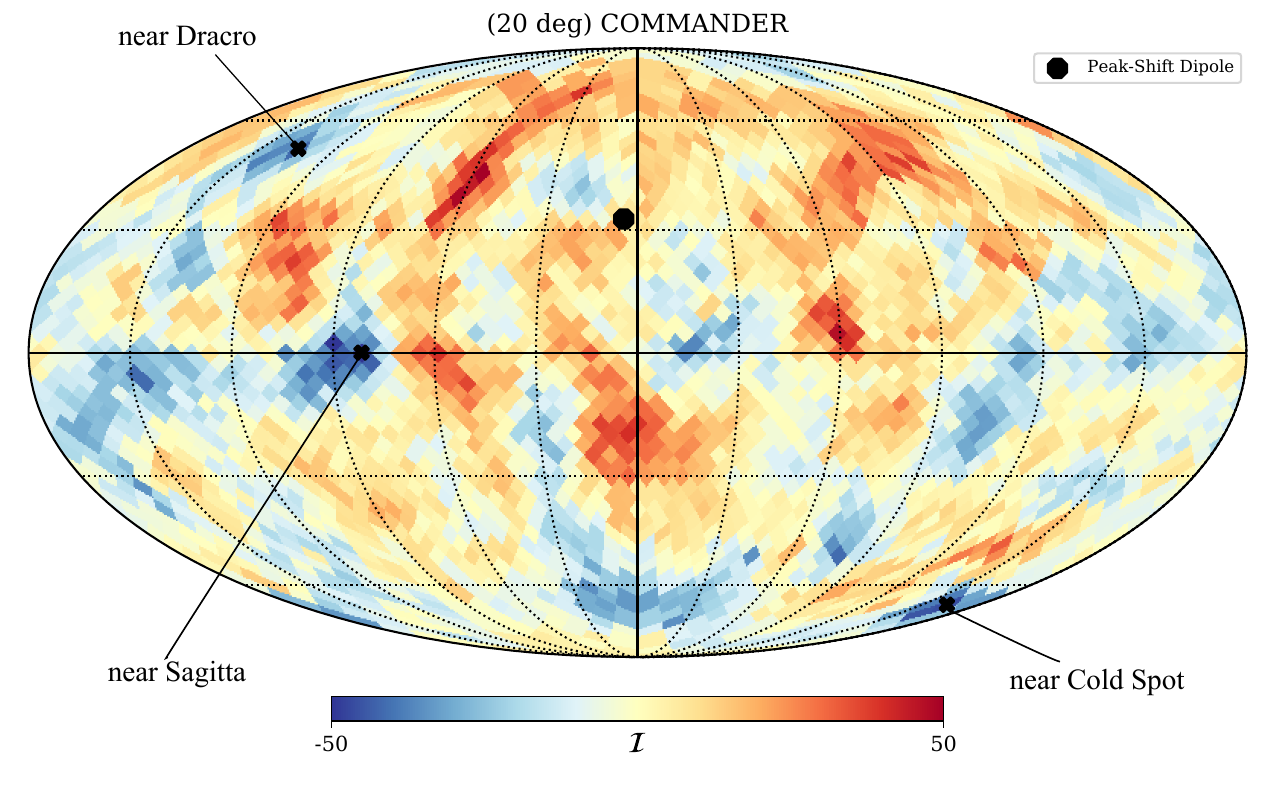}
    \includegraphics[width=0.45\columnwidth]{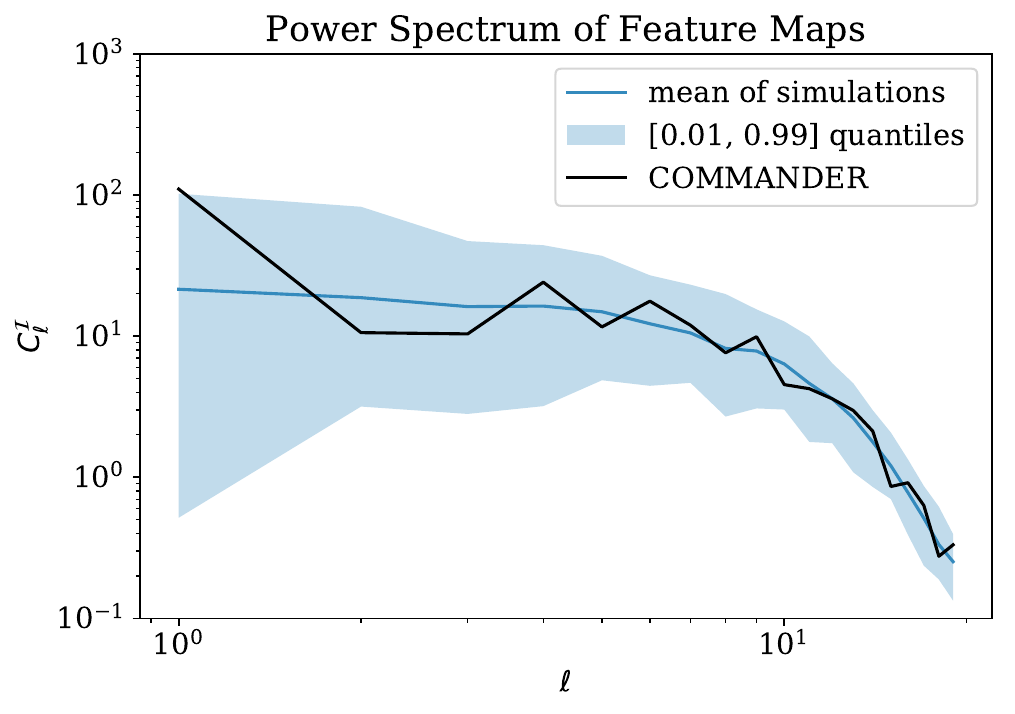}
    \includegraphics[width=0.45\columnwidth]{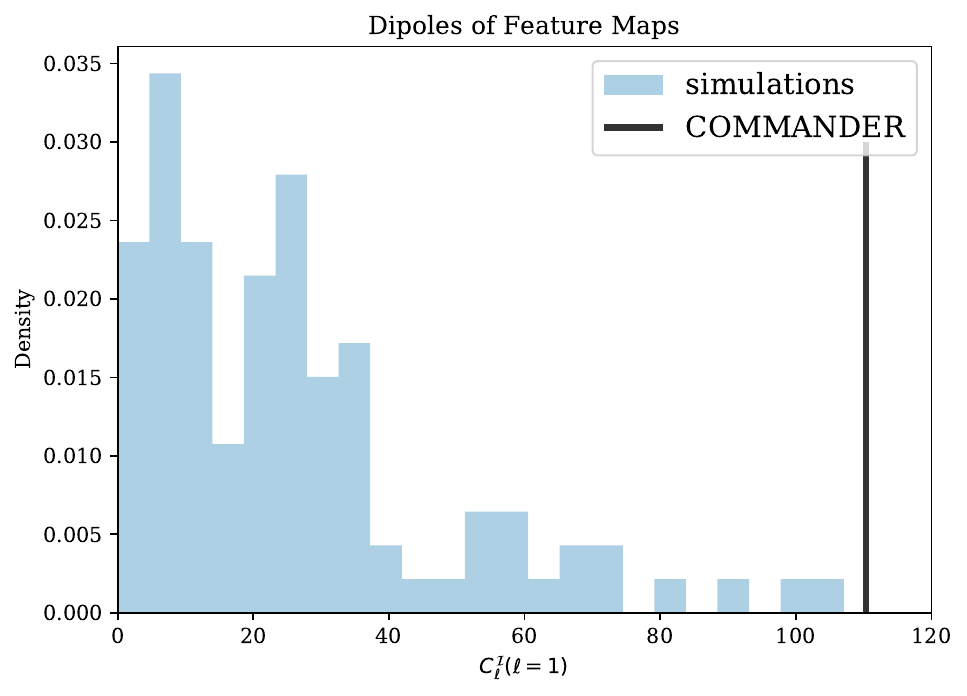}
    \caption{The full-sky peak-shift feature map in galactic coordinate and its power spectrum. We showed the full-sky peak shift in terms of $\mathcal{I}$ constructed from 30\,720 patches with $20\times20$ deg$^2$ size on \textit{Planck} COMMANDER map. The shift has a significantly strong dipole mode towards small scales (positive $\mathcal{I}$) at $(l,b)=(4^\circ.5, 32^\circ.7)$. In the bottom left we show the power spectrum $C^{\mathcal{I}}_\ell$ of this peak-shift $\mathcal{I}$ feature map (black curve) and those from 100 simulations (light blue area).}
    \label{fig:CS-Commander-Hyperlarge-I}
\end{figure}

\begin{figure}
    \centering
    \includegraphics[width=\columnwidth]{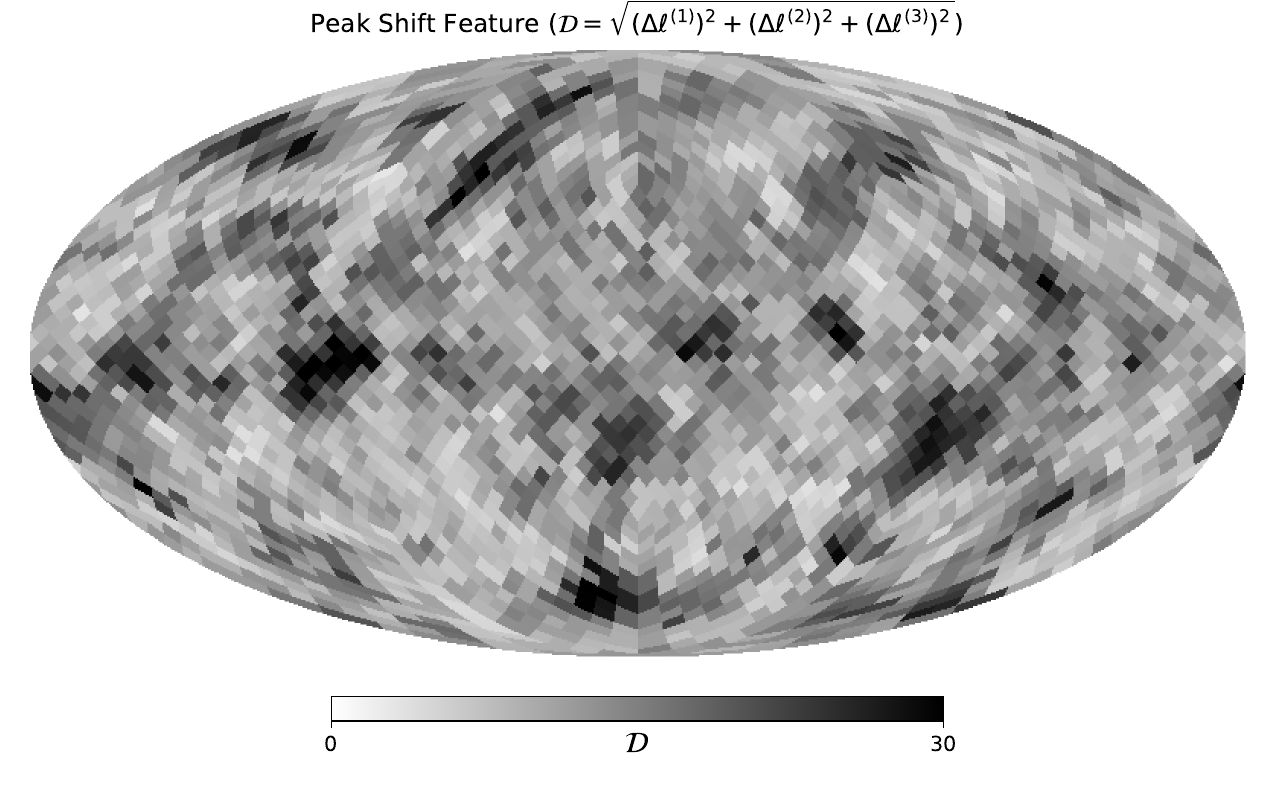}
    \caption{The full-sky map for the distance feature ($\mathcal{D} = \sqrt{(\Delta \ell^{(1)})^2 + (\Delta \ell^{(2)})^2 + (\Delta \ell^{(3)})^2}$).}
    \label{fig:CS-Commander-Hyperlarge-D}
\end{figure}

\begin{figure}
    \includegraphics[width=\columnwidth]{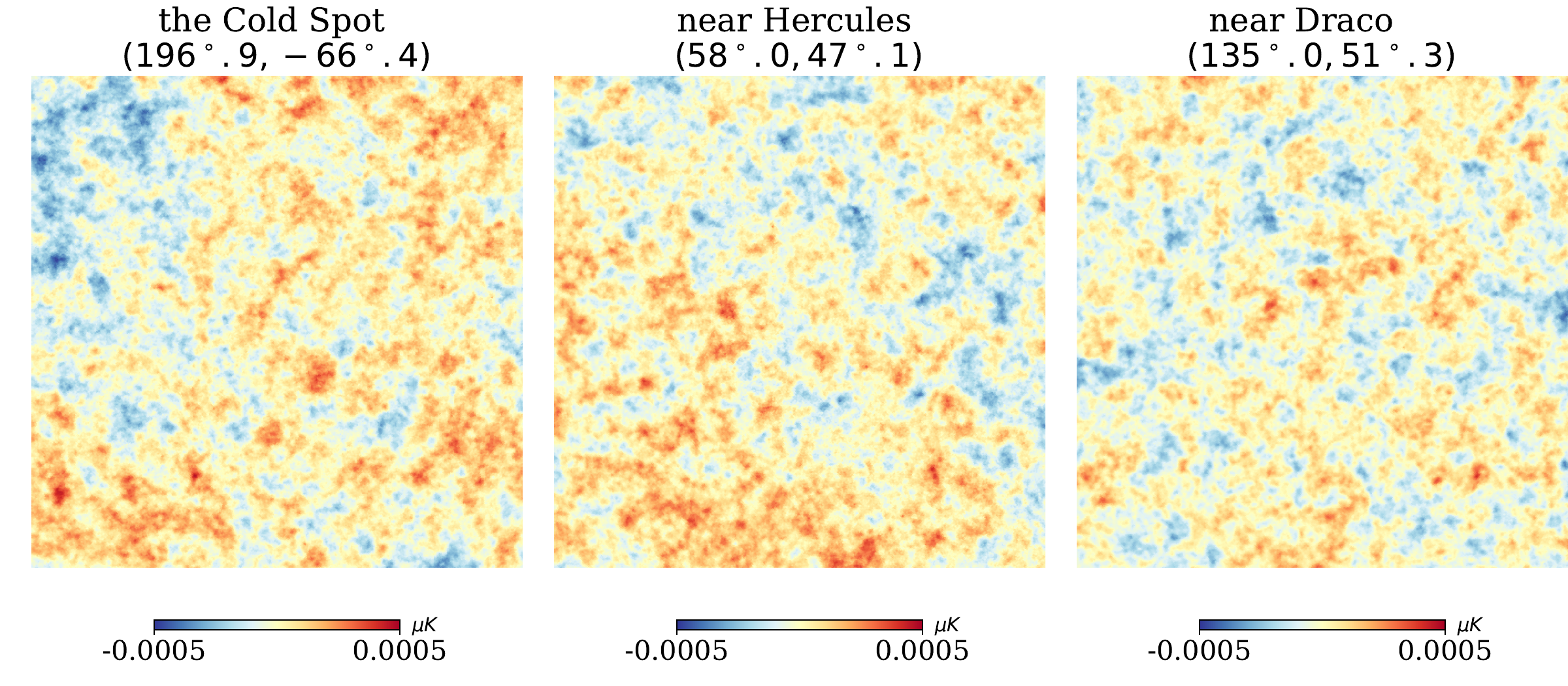}
    \includegraphics[width=\columnwidth]{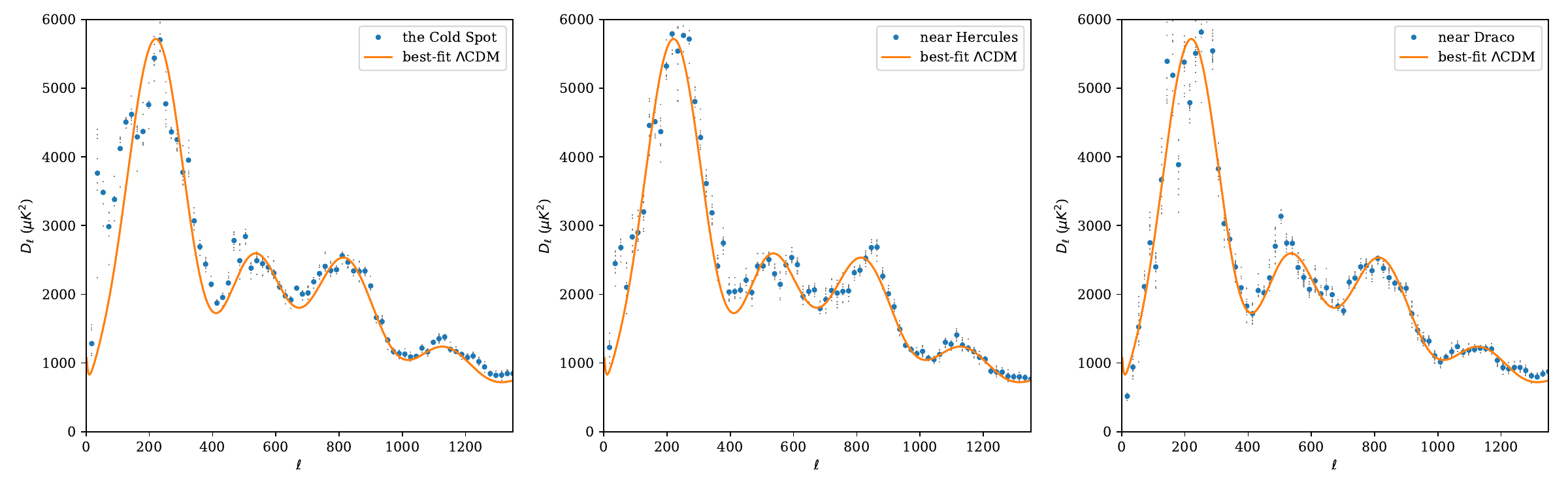}
    \caption{\textbf{Top panels:} Three selected CMB patches in three different pixels on the peak-shift feature map. From left to right patches are the patch at the Cold Spot pixel, the patch at the Hercules pixel, and the one at the Draco pixel.
    \textbf{Bottom panels:} Mean power spectra from three selected pixels: the Cold spot pixel (with total shift vale $\mathcal{I} = -45.93$), Hercules pixel (with total shift value $\mathcal{I} = + 50.09$), and Draco pixel (with total shift value $\mathcal{I} = -41.37$). Blue points are the mean power spectra of the selected pixels, black dots are all power spectra in the selected pixel, and the orange curve is the best-fit $\Lambda$CDM model from \textit{Planck}.
    }
    \label{fig:Selected-Patches}
\end{figure}

We show the peak-shift feature $\mathcal{I}$ map in Fig~\ref{fig:CS-Commander-Hyperlarge-I} and peak-shift feature $\mathcal{D}$ map in Fig~\ref{fig:CS-Commander-Hyperlarge-D}. 
One can see in Fig~\ref{fig:CS-Commander-Hyperlarge-I} the area near the Cold Spot, which was already reported in \cite{Chiang2018}, 
has a $\mathcal{I}=-43.6\pm6.7$ at $(l,b)\simeq(196^\circ.9,-66^\circ.4)$\footnote{Since we set our feature map on a $N_{\rm side}=16$ HEALPix grid, the resolution uncertainty of this position is around $3^\circ.7$.}.
The Cold Spot is the 2nd most negative (2nd largest peak shift towards large scales) in the feature map.
The one with the most negative shift is near the area of $(l,b)\simeq(81^\circ.6, 0^\circ.0)$ with $\mathcal{I} = -42.9\pm1.5$. The 3rd most negative is located near Draco supervoid \cite{Szapudi2014} at $(l,b)\simeq(135^\circ.0,51^\circ.3)$ with $\mathcal{I} = -41.37\pm2.6$. 
Also, see Fig~\ref{fig:Selected-Patches} for the CMB patches and power spectra of selected pixels.

The uncertainties are bootstrap resampled from the peak shift values within a given HEALPix pixel.
We resampled the $I$ values in each pixel with replacement, and estimated the uncertainties by,
\begin{equation}
    \sigma^2 = \frac{1}{M} \Sigma_{j = 1}^M [m_j - m],
\end{equation}
where $M$ is the number of bootstrap trails, $m_j$ is the the $I$ estimated value for a nearby patch, and $m$ we used the median value of all the $I$ values of a given pixel.

\begin{figure}
    \includegraphics[width=\columnwidth]{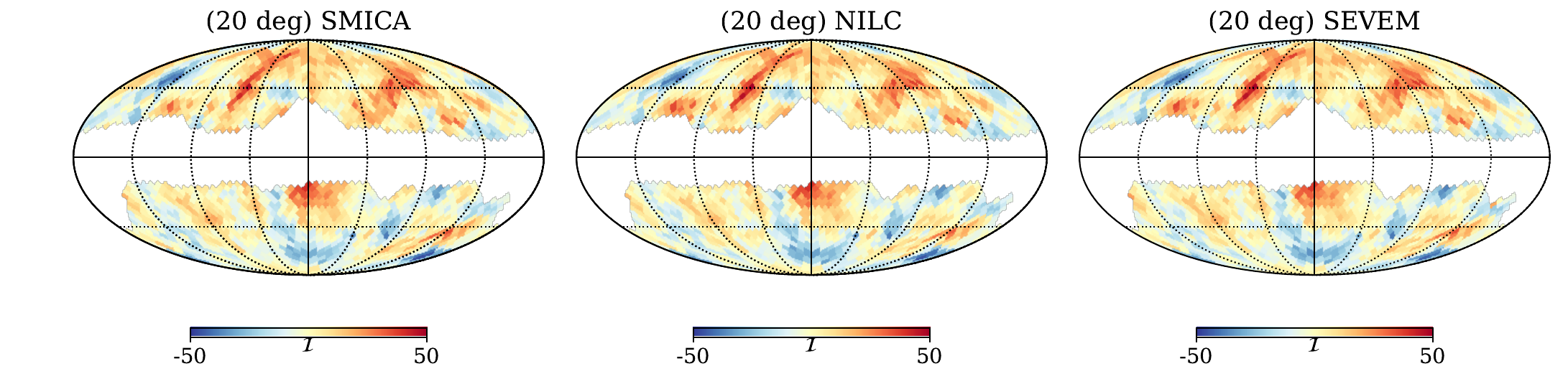}\\
    \includegraphics[width=\columnwidth]{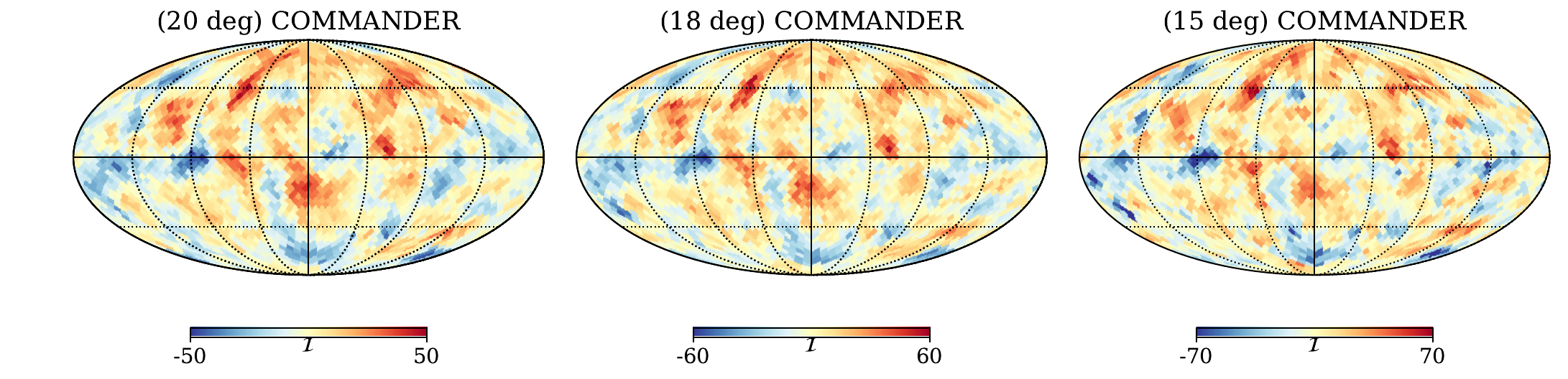}
    \caption{Feature $\mathcal{I}$ maps based on SMICA, NILC, SEVEM CMB temperature maps, with \textit{Planck} GAL70 mask. One can see the close resemblance of the large-scale features outside the mask the $\mathcal{I}$. In the bottom panels we show the COMMANDER $\mathcal{I}$ feature maps constructed with different sizes of patches, $20^2$, $18^2$ and $15^2$ deg$^2$, respectively. Each feature map is constructed with 30\,720 patches from the COMMANDER map on HEALPix $N_{\rm side}=16$. 
    }
    \label{fig:Feature-Subplots-Components-Sizes}
\end{figure}

\begin{figure}
    \centering
    \includegraphics[width=\columnwidth]{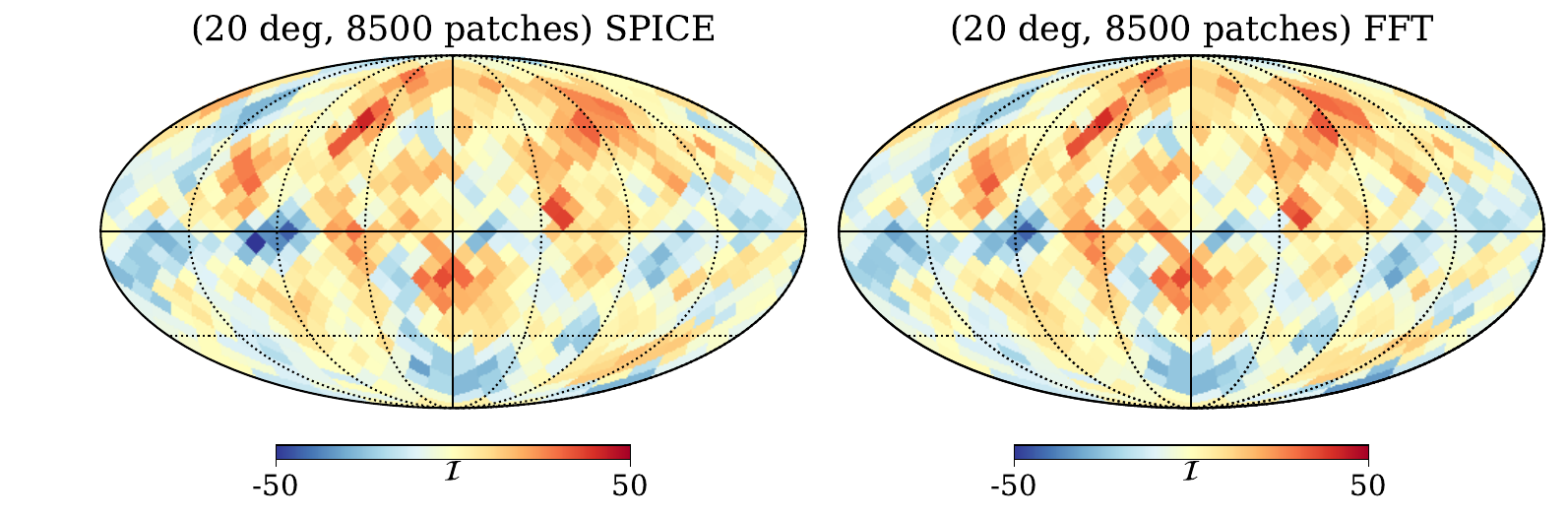}
    \caption{Comparison of $\mathcal{I}$ feature maps with SPICE (left) and FFT from flat sky approximation (right) power spectrum extraction method. We use 8\,500 patches of $20^2$ deg$^2$ on HEALPix $N_{\rm side}=8$.}
    \label{fig:Feature-Subplots-SPICE}
\end{figure}

For comparison, we also processed the other three \textit{Planck} maps SMICA, NILC, SEVEM using \textit{Planck} GAL70 mask in the top panels of Fig.~\ref{fig:Feature-Subplots-Components-Sizes}.
The bottom panels show the comparison of the feature maps with different patch sizes $15^2$, $18^2$, and $20^2$ deg$^2$. 
Another comparison was made on the feature map for power spectra taken from SPICE package \cite{Szapudi2001} and from FFT in Fig.~\ref{fig:Feature-Subplots-SPICE}. 
We used 8\,500 patches of $20^2$ deg$^2$ on HEALPix $N_{\rm side}=8$ for the comparison between FFT and SPICE, and we found no significant difference.
We, therefore, chose to use FFT for the reason of computational speed.

\section{Analysis of Results}
In analyzing the full-sky peak-shift feature map with spherical harmonic multipole expansion, 
we computed the angular power spectrum of the feature map in the bottom left of Fig.~\ref{fig:CS-Commander-Hyperlarge-I}. 
We found that the dipole $C^{\mathcal{I}}_{\ell=1}$ of the feature map was unusually large, with magnitude $\simeq 102.6\pm2.8$.

\begin{figure}
    \includegraphics[width=\columnwidth]{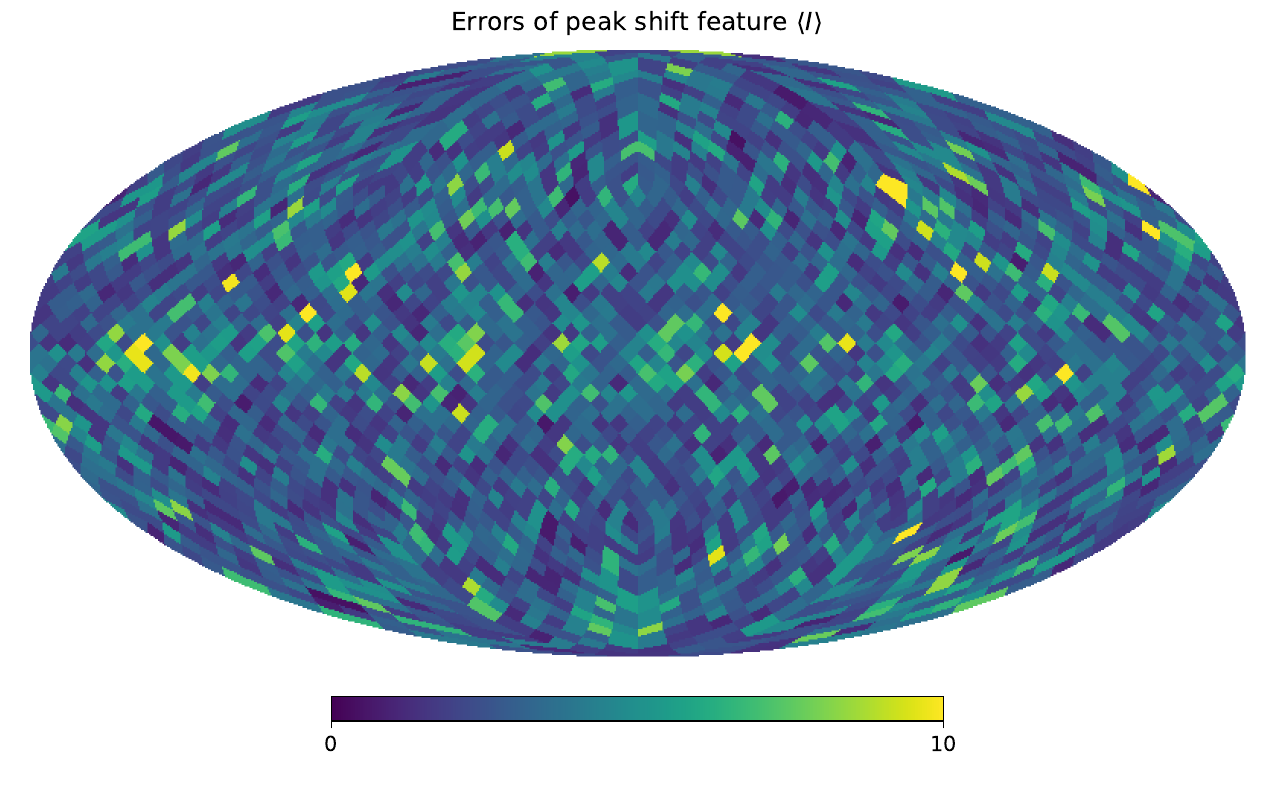}
    \caption{
        Feature errors of the Fig~\ref{fig:CS-Commander-Hyperlarge-I}. 
        The unit is in $\ell$, since $I = \ell^{(1)} + \ell^{(2)} + \ell^{(3)}$.}
    \label{fig:errors}
\end{figure}

The uncertainty in dipole mode was calculated by assuming each HEALPix pixel on the feature map is governed by a Gaussian distribution with its mean equals to the pixel value, and the standard deviation equals to the bootstrap error.
The value of dipole mode may vary according to the distribution of the pixel uncertainties on the full-sky.
We, therefore, sampled realizations of the feature map based on the bootstrap errors of pixels and computed the uncertainty of the dipole mode using the random realizations.
The error in each pixel represents how much the peak positions would be shifted according to the slight local movements of the $20^2$ deg$^2$ filter (patch).
The distribution of the feature errors is shown in Fig.~\ref{fig:errors}, which is roughly isotropic except the galactic plane area.

Two known sources can cause significant peak shift in small patches, i.e., gravitational lensing \cite{Carron2017}, and Doppler Boosting effect \cite{Planck2013xvii,Jeong2014}.
We took the \textit{Planck} 100 lensing simulations from the official website \cite{Planck2013xvii}
\footnote{\tt{https://wiki.cosmos.esa.int/planckpla/index.php/Simulation\_data/}} as a test set and ran the same feature-map-making procedure.
These \textit{Planck} simulations are based on the power spectrum of the best-fit $\Lambda$CDM model with gravitational lensing and Doppler boosting effect included. 
We constructed full-sky feature maps with the same method described above from those simulations. 
The dipole of the COMMANDER feature map was higher than those from all the 100 simulations with significance $p < 1\%$, as is shown in the bottom right panel of Fig.~\ref{fig:CS-Commander-Hyperlarge-I}. 

Since the \textit{Planck} maps still contain the Doppler Boosting effect, one can correct the peak positions of the patches from such effect (particularly the aberration effect) via $\Delta \ell/\ell = \beta \cos{\theta}$ \cite{Jeong2014}, where $\theta$ is the angle deviated from the velocity direction, and $\beta \equiv v/c\simeq 0.00123$ is a constant, and the peak-shift dipole direction, after such correction, the pointing of this dipole is now from $(l,b)=  (-175^\circ.5, -32^\circ.7)$ to $(4^\circ.5, 32^\circ.7)$ (negative to positive $I$).
    
To test the influences of foreground residuals on the peak-shift dipole, we added three different types of 5\% foregrounds: free-free, synchrotron, and the thermal dust maps to 100 CMB simulations. 
The 5\% amplitude of the foregrounds on the galactic plane region is still too high to be considered as foreground residuals, so we further block out the Gal90 galactic plane region to make the foregrounds to have a similar amplitude in the other areas. 
We used the foreground maps released by \cite{Planck2015x} and adjusted their reference frequency to be 100 GHz. 
The signal models for different CMB foreground brightness temperature were given in Table 4 of \cite{Planck2015x}. 
We processed the 100 CMB simulations with 5\% foregrounds (free-free, synchrotron, and dust) using the same peak-shift feature map generation pipeline. 
With these 100 feature map simulations, we calculated the dipole power from the power spectrum of each feature map simulations. 
  
\begin{figure}
    \centering
    \includegraphics[width=0.48\columnwidth]{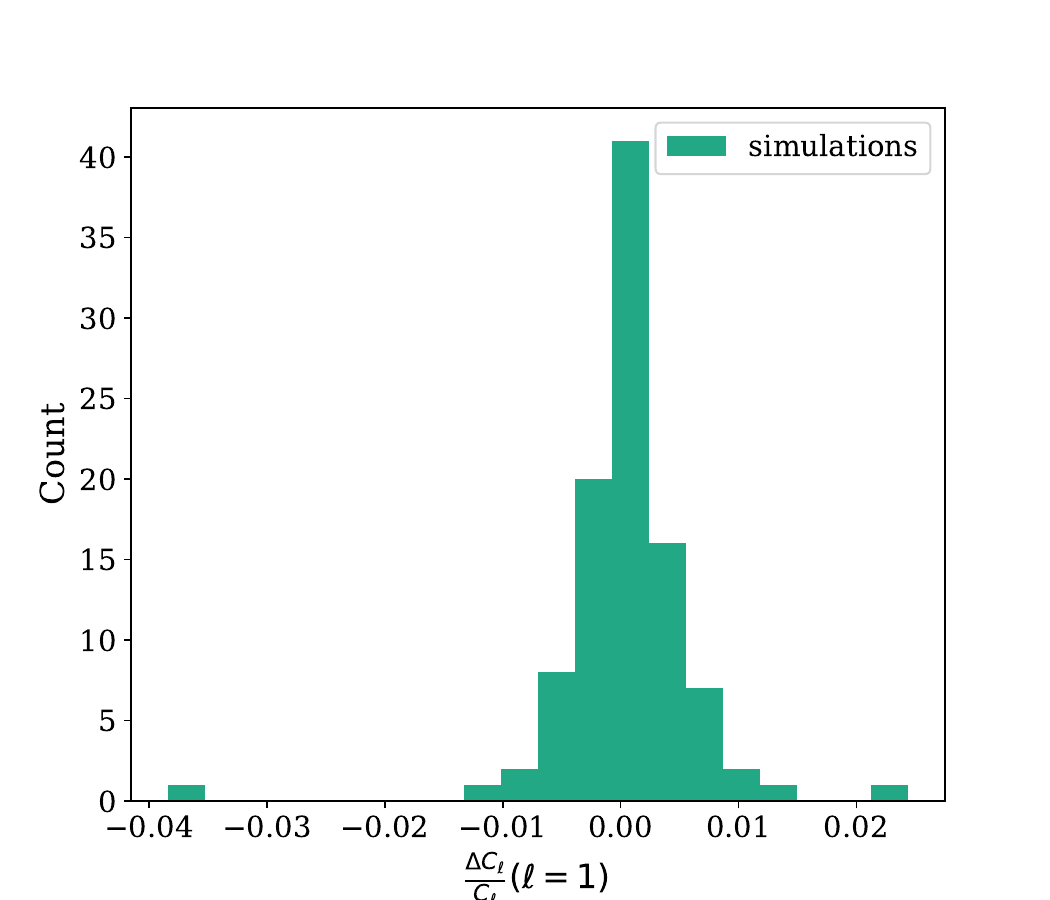}
    \includegraphics[width=0.48\columnwidth]{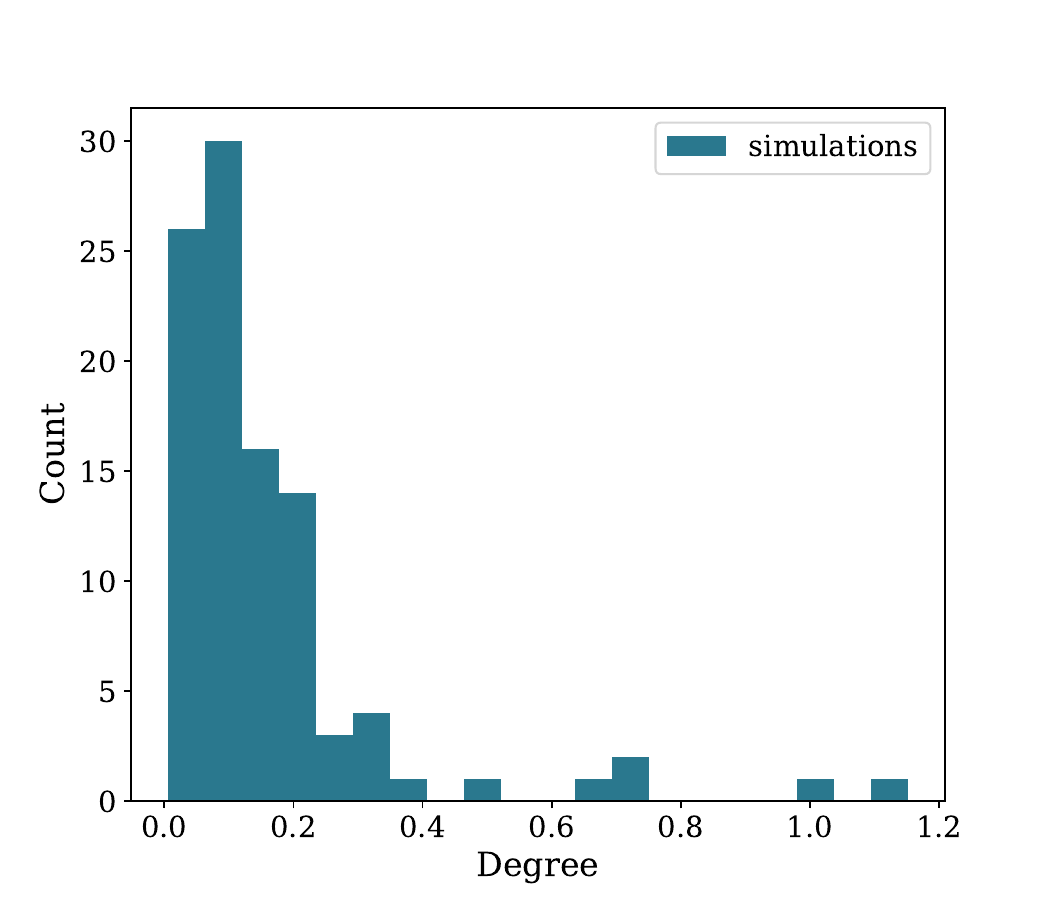}
    \caption{The effect of 5\% foregrounds on the direction and power of the peak-shift dipole. 
    We added 5\% \textit{Planck} foregrounds:free-free, thermal dust, synchrotron emission scaled to 100 GHz to 100 simulated maps and calculated the change in dipole power (left) and dipole direction (right). 
    The standard deviation of the change in dipole power is $0.61$\%, and the mean of change in direction is less than $0^\circ.16$, meaning the foreground residual cannot be the cause of the significantly strong dipole.}
    \label{fig:Dipole-Foreground-Full-Sky}
\end{figure}

In Fig.~\ref{fig:Dipole-Foreground-Full-Sky}, we plot the difference in the dipole power and the direction shift after adding the foregrounds. 
We found the change of the standard deviation of the dipole magnitude ($\Delta C^{\mathcal{I}}_{\ell=1}/C^{\mathcal{I}}_{\ell=1}$) is $0.61$\% and the mean shift in dipole direction is $0^\circ.16$ among the 100 simulations. 
Thus foreground residual has a small effect on the direction and magnitude of the dipole. 

\section{Discussion}

We offer a new way to visualize the statistics of spherical data using a feature map.
We find large peak shifts in the Cold Spot area and around the Draco constellation.
We also find there exists a peak-shift dipole in our feature map according to our analysis. 
Since we have tested the foreground contamination, another possible cause could be the dipoles related to motion with respect to the CMB restframe or the uncertainty related to our fitting method.
In our fitting method, though we consider the uncertainty of generating a feature map, we did not consider the peaks' uncertainties according to fitting Gaussian curves to the power spectrum.
The dipole on the feature map could be possibly resolved after considering the uncertainty of fitting Gaussian curves.
Further development of the method is needed to understand the underlying uncertainty of the fitting procedure.

This dipole related to the motion is usually subtracted from the CMB data during data processing. 
It is mentioned in \cite{Planck2018i} that the 2015 nominal Solar dipole is slightly different from the best Solar dipole in 2018 DR3 maps,
though the 2015 nominal Solar dipole was used for all 2018 maps.
Since we haven't taken into account systematics in motion in our analysis,
the dipole we found in the feature map is possible due to the systematics in the Solar dipole measurement or other systematics related to motion.
Further tests need to be performed to understand how the systematics of motion (with respect to the CMB restframe) affect our feature map.


\acknowledgments
The authors acknowledge the use of ESA \textit{Planck} Legacy Archive data\footnote{\tt{https://pla.esac.esa.int/\#home/}}
, HEALPix\footnote{\tt{http://healpix.jpl.nasa.gov/}} package \cite{Gorski2005} 
and GLESP\footnote{\tt{http://www.glesp.nbi.dk/}} 
package \cite{Doroshkevich2005}. 

\bibliographystyle{JHEP}
\bibliography{sample}

\providecommand{\href}[2]{#2}\begingroup\raggedright\begin{thebibliography}{10}

\bibitem{Hu2002}
W.~Hu and S.~Dodelson, \emph{Cosmic microwave background anisotropies},
  {\emph{Annual Review of Astronomy and Astrophysics} {\bfseries 40} (2002)
  171}.

\bibitem{Dodelson2003216}
S.~Dodelson, \emph{8 - anisotropies},  in \emph{Modern Cosmology}, pp.~216 --
  260, Academic Press, (2003).

\bibitem{wmap7yanomalies}
C.~Bennett, R.~Hill, G.~Hinshaw, D.~Larson, K.~Smith and J.~Dunkley,
  \emph{Seven-year wilkinson microwave anisotropy probe (wmap) observations:
  Are there cosmic microwave background anomalies}, {\emph{ApJ} {\bfseries 192}
  (2011) 17}.

\bibitem{Planck2015xvi}
{Planck Collaboration}, {Ade, P. A. R.}, {Aghanim, N.}, {Akrami, Y.}, {Aluri,
  P. K.}, {Arnaud, M.} et~al., \emph{Planck 2015 results - xvi. isotropy and
  statistics of the cmb}, {\emph{A\&A} {\bfseries 594} (2016) 15:1}.

\bibitem{Carron2017}
J.~Carron, A.~Lewis and A.~Challinor, \emph{Internal delensing of planck cmb
  temperature and polarization}, {\emph{JCAP} {\bfseries 2017} (2017) 035}.

\bibitem{Planck2013xvii}
{Planck Collaboration}, {Ade, P. A. R.}, {Aghanim, N.}, {Armitage-Caplan, C.},
  {Arnaud, M.}, {Ashdown, M.} et~al., \emph{Planck 2013 results. xvii.
  gravitational lensing by large-scale structure}, {\emph{A\&A} {\bfseries 571}
  (2014) A17}.

\bibitem{Lewis2006}
A.~Lewis and A.~Challinor, \emph{Weak gravitational lensing of the cmb},
  {\emph{Phys. Rep.} {\bfseries 429} (2006) 1 }.

\bibitem{Notari2014}
A.~Notari, M.~Quartin and R.~Catena, \emph{Cmb aberration and doppler effects
  as a source of hemispherical asymmetries}, {\emph{JCAP} {\bfseries 2014}
  (2014) 019}.

\bibitem{Chiang2018}
L.-Y. Chiang, \emph{Excessive shift of the cmb acoustic peaks of the cold spot
  area}, {\emph{ApJ} {\bfseries 861} (2018) 8}.

\bibitem{Vielva2004}
P.~Vielva, E.~Martínez-González, R.~B. Barreiro, J.~L. Sanz and L.~Cayón,
  \emph{Detection of non-gaussianity in the wilkinson microwave anisotropy
  probe first-year data using spherical wavelets}, {\emph{ApJ} {\bfseries 609}
  (2004) 22}.

\bibitem{Finelli2016}
F.~Finelli, J.~García-Bellido, A.~Kovács, F.~Paci and I.~Szapudi,
  \emph{Supervoids in the wise–2mass catalogue imprinting cold spots in the
  cosmic microwave background}, {\emph{MNRAS} {\bfseries 455} (2016) 1246}.

\bibitem{Planck2015ix}
{Planck Collaboration}, {Adam, R.}, {Ade, P. A. R.}, {Aghanim, N.}, {Arnaud,
  M.}, {Ashdown, M.} et~al., \emph{Planck 2015 results - ix. diffuse component
  separation: Cmb maps}, {\emph{A\&A} {\bfseries 594} (2016) A9}.

\bibitem{Chiang2012}
L.-Y. Chiang and F.-F. Chen, \emph{Direct measurement of the angular power
  spectrum of cosmic microwave background temperature anisotropies in the wmap
  data}, {\emph{ApJ} {\bfseries 751} (2012) 43}.

\bibitem{Chiang2011}
L.-Y. Chiang and F.-F. Chen, \emph{Cross-power spectrum and its application on
  window functions in the wilkinson microwave anisotropy probe data},
  {\emph{ApJ} {\bfseries 738} (2011) 188}.

\bibitem{More1977}
J.~More, \emph{Levenberg--marquardt algorithm: implementation and theory},
  {\emph{Conference: Conference on numerical analysis} (1977) }.

\bibitem{Page2003}
L.~Page, M.~R. Nolta, C.~Barnes, C.~L. Bennett, M.~Halpern, G.~Hinshaw et~al.,
  \emph{First-year wilkinson microwave anisotropy probe (wmap) observations:
  Interpretation of the tt and te angular power spectrum peaks}, {\emph{ApJS}
  {\bfseries 148} (2003) 233}.

\bibitem{Planck2015xi}
{Planck Collaboration}, {Aghanim, N.}, {Arnaud, M.}, {Ashdown, M.}, {Aumont,
  J.}, {Baccigalupi, C.} et~al., \emph{Planck 2015 results - xi. cmb power
  spectra, likelihoods, and robustness of parameters}, {\emph{A\&A} {\bfseries
  594} (2016) A11}.

\bibitem{Zhu1997}
C.~Zhu, R.~H. Byrd, P.~Lu and J.~Nocedal, \emph{Algorithm 778: L-bfgs-b:
  Fortran subroutines for large-scale bound-constrained optimization},
  {\emph{ACM Trans. Math. Softw.} {\bfseries 23} (1997) 550}.

\bibitem{Powell1964}
M.~J.~D. Powell, \emph{An efficient method for finding the minimum of a
  function of several variables without calculating derivatives}, {\emph{The
  Computer Journal} {\bfseries 7} (1964) 155}.

\bibitem{Nash1984}
S.~G. Nash, \emph{Newton-type minimization via the lanczos method}, {\emph{SIAM
  Journal on Numerical Analysis} {\bfseries 21} (1984) 770}.

\bibitem{Powell1998}
M.~J.~D. Powell, \emph{Direct search algorithms for optimization calculations},
  {\emph{Acta Numerica} (1998) 287–336}.

\bibitem{Nelder1965}
M.~J.~D. Powell, \emph{A simplex method for function minimization}, {\emph{The
  Computer Journal} {\bfseries 7} (1965) 308}.

\bibitem{Gorski2005}
K.~M. {G{\'o}rski}, E.~{Hivon}, A.~J. {Banday}, B.~D. {Wandelt}, F.~K.
  {Hansen}, M.~{Reinecke} et~al., \emph{Healpix: A framework for
  high-resolution discretization and fast analysis of data distributed on the
  sphere}, {\emph{ApJ} {\bfseries 622} (2005) 759}.

\bibitem{Kovacs2015}
A.~Kov\a'acs and I.~Szapudi, \emph{Star–galaxy separation strategies for
  wise-2mass all-sky infrared galaxy catalogues}, {\emph{MNRAS} {\bfseries 448}
  (2015) 1305}.

\bibitem{Szapudi2014}
I.~e.~a. Szapudi, \emph{{The Cold Spot in the Cosmic Microwave Background: the
  Shadow of a Supervoid}},  in \emph{{Proceedings, 49th Rencontres de Moriond
  on Cosmology: La Thuile, Italy, March 15-22, 2014}}, pp.~33--41, 2014,
  \href{https://arxiv.org/abs/1406.3622}{{\ttfamily 1406.3622}}.

\bibitem{Szapudi2001}
I.~Szapudi, S.~Prunet, D.~Pogosyan, A.~S. Szalay and J.~R. Bond, \emph{Fast
  cosmic microwave background analyses via correlation functions}, {\emph{ApJ}
  {\bfseries 548} (2001) L115}.

\bibitem{Jeong2014}
D.~Jeong, J.~Chluba, L.~Dai, M.~Kamionkowski and X.~Wang, \emph{Effect of
  aberration on partial-sky measurements of the cosmic microwave background
  temperature power spectrum}, {\emph{Phys. Rev. D} {\bfseries 89} (2014)
  023003}.

\bibitem{Planck2015x}
{Planck Collaboration}, {Adam, R.}, {Ade, P. A. R.}, {Aghanim, N.}, {Alves, M.
  I. R.}, {Arnaud, M.} et~al., \emph{Planck 2015 results - x. diffuse component
  separation: Foreground maps}, {\emph{A\&A} {\bfseries 594} (2016) A10}.

\bibitem{Planck2018i}
{Planck Collaboration}, Y.~{Akrami}, F.~{Arroja}, M.~{Ashdown}, J.~{Aumont},
  C.~{Baccigalupi} et~al., \emph{{Planck 2018 results. I. Overview and the
  cosmological legacy of Planck}}, {\emph{arXiv e-prints} (2018)
  arXiv:1807.06205} [\href{https://arxiv.org/abs/1807.06205}{{\ttfamily
  1807.06205}}].

\bibitem{Doroshkevich2005}
A.~G. Doroshkevich, P.~D. Naselsky, O.~V. Verkhodanov, D.~I. Novikov, V.~I.
  Turchaninov, I.~D. Novikov et~al., \emph{Gauss–legendre sky pixelization
  (glesp) for cmb maps}, {\emph{Int. J. of Mod. Phys. D} {\bfseries 14} (2005)
  275}.

\end{thebibliography}\endgroup

\end{document}